\documentstyle[aps,multicol,epsf,amsmath,amssymb]{revtex}

\def\gs{\gtrsim}

\def\be{\begin{equation}}
\def\en{\end{equation}}                  
\newcommand{\bi}[1]{\mbox{\boldmath$#1$}}

\begin{document}
\draft
\bibliographystyle{prsty}
\title{
Glass Transition of Hard Sphere Systems: Molecular Dynamics and
Density Functional Theory
}
\author{
Kang Kim
and 
Toyonori Munakata
}
\address{
Department of Applied Mathematics and Physics, 
Graduate School of Informatics,\\
Kyoto University, Kyoto 606-8501, Japan
}
\date{\today}
\maketitle

\begin{abstract}
The glass transition of a hard sphere system is investigated within the 
framework of 
the density functional theory (DFT).
Molecular dynamics (MD) simulations are performed to study dynamical 
behavior of the system on the one hand and to provide 
the data to produce the density field for the DFT on the other hand.
Energy landscape analysis based on the DFT shows that there appears 
a metastable (local) free energy minimum  representing 
an amorphous state as the density is increased.
This state turns out to become stable, compared with the uniform
liquid, at some density, around which  we also observe sharp slowing
down of the $\alpha$ relaxation in MD simulations.
\end{abstract}


\begin{multicols}{2}
\newpage

Understanding of the universal mechanism of the glass transition
is one of the major challenges in the current condensed matter physics.
From a dynamical viewpoint, we would like to know how 
a drastic slowing down near the transition point (temperature or density) 
and an eventual exceeding of the relaxation time
over the experimental time scale are realized~\cite{Jackle,Ediger}.
Energetically or statically, it is asked whether a thermodynamic
glassy state with the amorphous arrangement of particles has lower
free energy than a liquid state of uniform density.  
In order to answer these questions,
many efforts have been devoted to real experiments, computer
simulations, and theories for the last decades.

As one of theories to study supercooled liquids and
glasses, the density functional theory (DFT) is recently gathering 
much attention~\cite{DFT}.
The DFT is now a conventional method to study the
freezing~\cite{Ramakrishnan,Haymet} and other transitions.  
The glass transition has been investigated also based on 
the DFT by some workers~\cite{Singh,Baus,Lowen,Das,Das2}.
In the earlier works~\cite{Singh,Baus,Lowen,Das,Das2}, the random
close packing (RCP) of hard spheres has been produced by the Bennett's
algorithm~\cite{Bennett} and the free energy from the DFT with the
input density field supplied by the RCP data has been calculated.
Singh {\it et al.}~\cite{Singh} showed that the hard sphere glassy
state becomes more stable than a uniform liquid at the critical
density $n_{g}\sigma^3=1.14$, suggesting that there exists
a kind of the thermodynamic (later called a random first
order~\cite{Xia}) glass transition.
Here $\sigma$ is the hard sphere diameter and $n$ is the number
density of the system.
It is remarked here that since the RCP configurations were produced 
by a kind of aggregation method, we can not study dynamical aspects
of the glass transition found by the energetics based on DFT.

The purpose of this paper is first to produce a supercooled and a 
glassy state for a one-component hard sphere system, relying  
on the uniform compression molecular dynamics (MD) method recently
developed by Lubachevsky and Stillinger~\cite{Stillinger}
and then to study both a dynamical and static properties of the state.
Especially from the particle configuration data, we can discuss free
energy within the DFT framework. 
This MD approach, in conjunction with the DFT, enables us to study 
both dynamical and static aspects of the transition in contrast 
with the Bennett's approach.      

Our system consists of $N=1372$ identical hard spheres with the mass
$m$ and the diameter $\sigma$
in a cubic box of volume $V$ with periodic boundary
conditions.
Throughout this paper, the units of length and time are
$\sigma$ and $\sqrt{{m\sigma^2}/{k_BT}}$, respectively, where $k_B$ is
Boltzmann's constant and $T$ is the temperature~\cite{units}.
It should be noted that the temperature is fixed as $k_BT=1$ in the
course of MD simulations.

To begin with, we briefly explain our MD method to obtain glassy
states of a one-component hard sphere system.
Employing the standard Alder and Wainwright
algorithm~\cite{Alder,Allen},
we first generate the equilibrium liquid state at the density
$\tilde n=0.86$.
It is well known that the fluid system freezes at $\tilde n_f\simeq
0.94$.
To avoid the crystallization and obtain amorphous glassy states,
Lubachevsky and Stillinger introduced a compressing (or quenching)
procedure~\cite{Stillinger}, in which 
they actually 
increased the diameter $\sigma$ with
a constant rate of expansion during MD simulations.
The dimensionless expanding rate $\Gamma$ is defined as
\begin{equation}
\Gamma=\frac{d\sigma(t)}{dt}\sqrt{\frac{m}{k_BT}},
\end{equation}
and $\Gamma=0.01$ is chosen in our simulations.
From the initial state $\tilde n=0.86$, we expanded each sphere with
the rate $\Gamma$ and 
could obtain various high density states, $0.86$,
$0.94$, $1.02$, $1.06$, $1.10$, $1.14$, $1.18$ and $1.21$, without
crystallization.

\begin{figure}
\epsfxsize=3.in
\centerline{\epsfbox{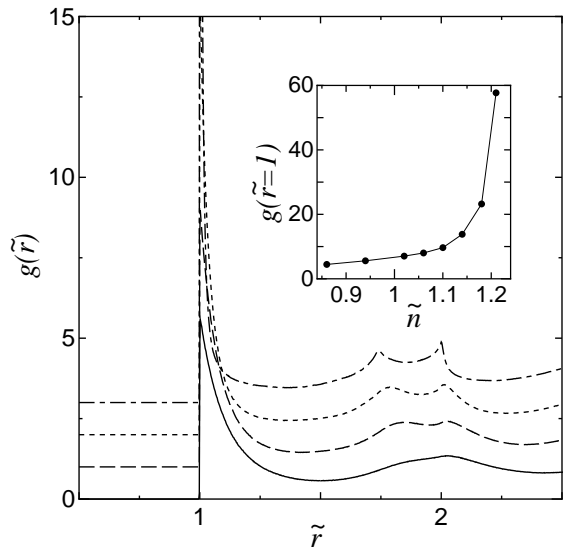}}
\caption{
The radial distribution function $g(r)$ 
obtained for $\tilde n=0.94$ (solid line), $1.06$ (dashed line), $1.14$
(short dashed line), and $1.21$ (dot dashed line).
Inset: contact value of radial distribution function $g(r=\sigma)$ as a
function of $\tilde n$.
The units of $\tilde r$ and $\tilde n$ are
$\sigma$ ($\tilde r=r\sigma$) and $\sigma^{-3}$ ($\tilde n=n\sigma^3$),
respectively.
}
\label{gr}
\end{figure}

Let us first study the static structure of the system.
For the purpose the radial distribution function $g(r)$, which is defined
by
\begin{equation}
g(r)=\frac{1}{n N}
\Biggl\langle \sum_{i \ne j}^{N}\delta({\bi r}+{\bi r_i}-{\bi r_j})
\Biggr\rangle,
\end{equation}
is calculated, where ${\bi r_i}$  represents the 
positions of the $i$-th
particle  and $\langle\cdots\rangle$
denotes the ensemble average over different configurations.
In Fig.~\ref{gr}, we plotted $g(r)$ for the density $\tilde n=0.94$,
$1.06$, $1.14$, and $1.21$.
It is noted that there is no sign of 
crystallization, which would be reflected in sharp peaks of $g(r)$ at
some characteristic lattice spacings.
Instead, $g(r)$ for higher density ($\tilde n \ge 1.14$) shows the 
splitting of the second
peak, which is a familiar characteristic of a glassy state of a simple
liquid.
Furthermore, we notice that the contact value $g(r=\sigma)$ shows an
anomalous increase as the density increases (see the inset), 
which corresponds to the fact that the pressure increases drastically
with increasing $\tilde n$.
Incidentally it should be remarked that the forms of $g(r)$ agree
qualitatively with those illustrated in Ref.~\cite{Bennett}.

\begin{figure}
\epsfxsize=3.in
\centerline{\epsfbox{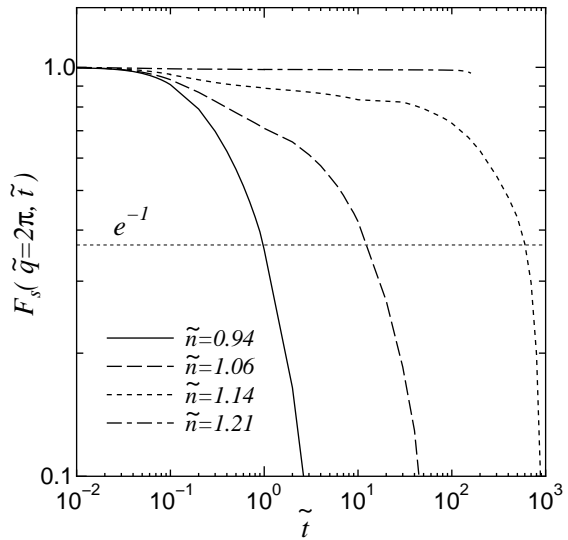}}
\caption{
Intermediate scattering function $F_s(q,t)$ at a wave
number $\tilde q=2\pi$
for
$\tilde n=0.94$ (solid line), $1.06$ (dashed line), $1.14$
(short dashed line), and $1.21$ (dot dashed line).
The units of $q$ and $t$ are $\sigma$ ($\tilde q=q\sigma$) and
$\sqrt{{m\sigma^2}/{k_BT}}$ ($\tilde t=t\sqrt{{k_BT}/{m\sigma^2}}$),
respectively.
}
\label{fs}
\end{figure}

\begin{figure}
\epsfxsize=3.in
\centerline{\epsfbox{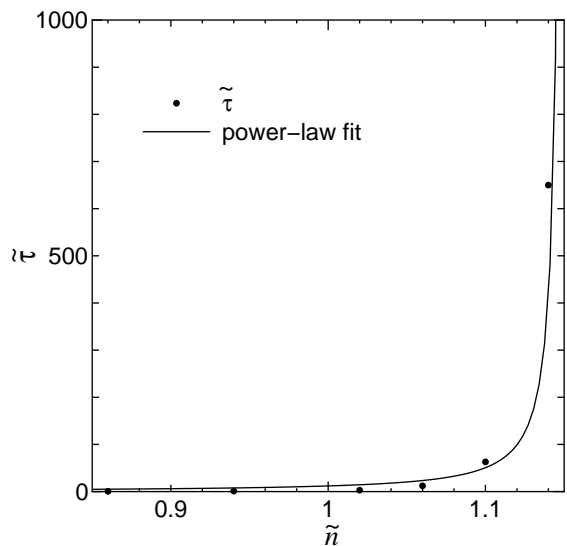}}
\caption{
Structural relaxation time $\tilde \tau$ as a function of density $\tilde n$
(closed circles).
Solid line represents power-law fit $(\tilde n_{g,MD}-\tilde
n)^{-\gamma}$ with $\tilde n_{g,MD}=1.15$ and $\gamma=1.31$.
}
\label{tau}
\end{figure}

We next consider the dynamical aspects of the hard sphere
glasses by calculating the incoherent intermediate scattering function
$F_s(q,t)$, which is defined by
\begin{equation}
F_s(q,t)=\Biggl\langle\frac{1}{N}\sum_{j=1}^{N} \exp[i{\bi q} \cdot
\Delta{\bi r}_j(t;t_0)]\Biggr\rangle_{t_0},
\end{equation}
where
$\Delta{\bi r}_j(t;t_0)={\bi r}_j(t+t_0)-{\bi r}_j(t_0)$ is 
the displacement vector of the $j$-th particle in time $t$, and
$\langle\cdots\rangle_{t_0}$ presents an average over initial times
$t_0$.
It is noted that the $F_s(q,t)$ is one of the standard quantities 
in studies of dynamical
properties of supercooled liquids and glasses~\cite{Kob,Yamamoto}.
In Fig.~\ref{fs}, we plotted the decay profiles of $F_s(q,t)$ at a
dimensionless wave number $\tilde q=2\pi$
for $\tilde n=0.94$, $1.06$, $1.14$, and $1.21$.
We see in Fig.~\ref{fs} that the relaxation of the $F_s(q,t)$ at
$\tilde n=0.94$ can be expressed by a simple exponential function.
Beyond the density $\tilde n=1.06$, however, $F_s(q,t)$ exhibits the
two-step, that is fast $\beta$ and slow $\alpha$, relaxation, which is
often mentioned as a characteristic sign of the slow relaxation in
glass forming liquids.
At the highest density $\tilde n=1.21$, the $F_s(q,t)$ dose not
show any decaying behavior~\cite{numerical}.
One can define the structural relaxation time $\tilde \tau$ by
$F_s(\tilde q=2\pi,\tilde \tau)=e^{-1}$ and this $\tilde \tau$ is plotted
as a function of $\tilde n$ in Fig.~\ref{tau}.
We notice in Fig.~\ref{tau} that the relaxation time $\tilde \tau$ shows a
strong dependence on the density, which can be expressed
by the power-law $(\tilde n_{g,MD}-\tilde n)^{-\gamma}$ with
$\tilde n_{g,MD}\simeq 1.15$ and $\gamma\simeq 1.31$
(solid line in Fig.~\ref{tau}).

We now consider energetics of the system based on the DFT and the 
configuration generated by the MD simulations.
For a practical calculation of the DFT, we employ the
Ramakrishnan and Yussouff (RY) free energy
functional~\cite{Ramakrishnan} because of its simplicity
and physical clarity.
The RY functional $F[n]$ is given by
\begin{eqnarray}
F[n]=F_{id}+F^{(2)}_{int}+F_{uni},
\end{eqnarray}
where
\begin{eqnarray}
F_{id}
= k_B T\int n({\bi r})\ln \biggl[\frac{n({\bi r})}{n}\biggr]d{\bi
  r},
\label{entropy}
\end{eqnarray}
\begin{eqnarray}
F^{(2)}_{int}=-\frac{1}{2}k_B T\int\int [n({\bi
  r})&-&n]C(|{\bi r}-{\bi r}'|) \nonumber\\
&\times& [n({\bi r}')-n] d{\bi r} d{\bi r}'.
\label{interaction1}
\end{eqnarray}
Here, $F_{id}$ and $F_{uni}$ represent
the ideal gas contribution and the excess free energy of the
uniform liquid state $n({\bi r})=n$, respectively.
$F^{(2)}_{int}$ represents  the second order term in the expansion 
around the uniform liquid state, thus all terms higher than  second 
neglected. 
We note that
$C({\bi r})$ is the direct correlation function of the uniform
liquid with  the density $n$~\cite{Hansen}.

In order to evaluate the free energy of the system, we need the
trial density field $n({\bi r})$, for which 
we employ a conventional Gaussian
superposition~\cite{Singh,Baus,Lowen,Das,Das2}.
That is, the density field $n({\bi r})$ is expressed 
by a sum of Gaussians with the centers
located at $N$  sites $\{{\bi r_i}\}$, which are given by our MD
simulation.
\begin{eqnarray}
n({\bi r})
= \sum_{i=1}^{N}
\biggl(\frac{\alpha}{\pi}\biggr)^{3/2}\exp[-\alpha({\bi r}-{\bi
  r_i}^2)]
\equiv \sum_{i=1}^N z({\bi r};{\bi r_i}),
\label{density}
\end{eqnarray}
where $\alpha^{-1}$, a variational parameter for the calculation of 
the free energy, is proportional to the mean square 
displacement of each particle.
Small (large) $\alpha$ represents the uniform 
liquid (localized amorphous) state.

When $\alpha$ is very large,  $F_{id}$ is asymptotically represented
by~\cite{Singh}
\begin{eqnarray}
F_{id}(\alpha)
\sim Nk_BT
\biggl[\biggl\{ \frac{3}{2} \ln\biggl(\frac{\alpha}{\pi}\biggr)- 
\frac{3}{2}\biggr\}-\ln n \biggr].
\label{log}
\end{eqnarray}
For a small  $\alpha$ region, we
calculated the integral Eq.~(\ref{entropy}) numerically.
We confirmed  that $F_{id}$ approaches 0 when $\tilde \alpha \to 0$ and 
noticed that $F_{id}$ coincides with the value
of Eq.~(\ref{log}) for $\tilde \alpha \gs 20$.

It is easy to see that the interaction term $F_{int}$ 
can be divided into three parts as~\cite{Singh}
\begin{eqnarray}
F^{(2)}_{int}(\alpha)
=-\frac{1}{2}Nk_BT \biggl\{
F_{int,1}(\alpha)&+&F_{int,2}(\alpha)\nonumber\\
&-&n\int C(r)d{\bi r}\biggr\}
\label{interaction3},
\end{eqnarray}
where $F_{int,1}(\alpha)$ represents the
self-interaction of a single Gaussian,
\begin{eqnarray}
F_{int,1}(\alpha)
=\int\int z({\bi r};{\bi 0})C(|{\bi r}-{\bi r'}|)z({\bi
  r'};{\bi 0})d{\bi r}d{\bi r'},
\end{eqnarray}
and $F_{int,2}(\alpha)$ represents the interaction 
between the two distinct Gaussians,
\begin{eqnarray}
F_{int,2}(\alpha)
=n\int g(r_1)d{\bi r_1}\int\int z({\bi r};{\bi
  0})C(|{\bi r}-{\bi r'}|)\nonumber\\
\times z({\bi r'};{\bi r_1})d{\bi r}d{\bi r'}
\label{fint2}.
\end{eqnarray}
The pair distribution
function $g(r)$ in this equation is determined from the MD simulation.
With respect to the direct correlation function $C(r)$,
we use Henderson-Grundke expression for $C(r)$,
which is known to be reliable, though empirical, even for high density 
hard sphere liquid~\cite{Henderson}.

\begin{figure}
\epsfxsize=3.in
\centerline{\epsfbox{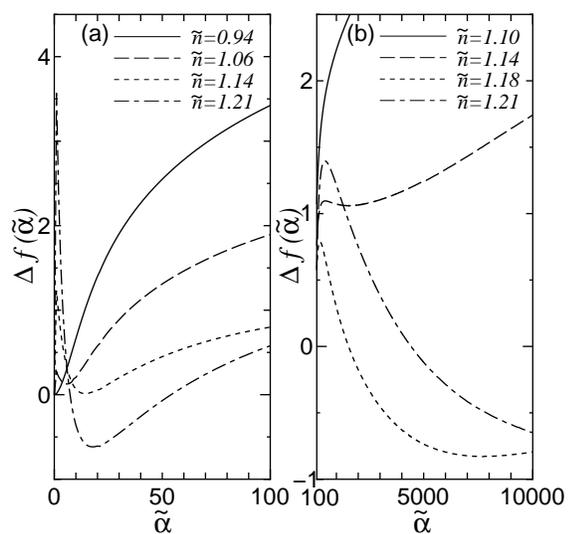}}
\caption{
Total free energy per particle relative to uniform liquid
$\Delta f(\tilde \alpha)$ for $\tilde \alpha \le 100$ (a) and $\tilde \alpha \ge 100$ (b)
as a function of localization parameter $\tilde \alpha$.
The unit of $\tilde \alpha$ is
 $\sigma^{-2}$ ($\tilde \alpha=\alpha\sigma^2$).
}
\label{df}
\end{figure}

The total free energy per particle relative to uniform state,
\begin{eqnarray}
\Delta f(\alpha)=
\frac{F_{id}(\alpha)+F^{(2)}_{int}(\alpha)}{Nk_BT},
\end{eqnarray}
is calculated
as a function of the localization parameter $\alpha$ (see
Figs.~\ref{df}).
It is seen in Figs.~\ref{df} that the free energy local minimum at
finite $\alpha$, which represents an amorphous state,
appears as the density is increased.
As is mentioned in Ref.~\cite{Singh}, the local
minimum appears as the result of the competition between the
ideal gas (simple increasing function of $\alpha$) and the interaction
terms.
Figures~\ref{df} show that two local minima are located at
$\tilde \alpha \simeq 13$ and $1600$ for $\tilde n=1.14$.
Das {\it et al.} also observe two local minima of $\Delta f(\alpha)$,
which are called the weakly localized state for small
$\alpha$ and the highly localized state for large $\alpha$,
respectively~\cite{Das}.
In addition, similar values for $\alpha$ 
are reported in Refs.~\cite{Singh,Das}, which also use the RY functional,
in relation to the local minimum of $\Delta f(\alpha)$.
As is stated in Ref.~\cite{Das}, the qualitative adequacy
is ambiguous for the highly localized state with very high value of
$\alpha$ since the RY form includes a perturbative
expansion around the uniform state.
In fact, based on the MD data we estimated $\alpha$ for 
high density states by relating it to the plateau value of the 
time dependent mean square displacement of 
each particle, yielding $\alpha \simeq 50$ for $\tilde
n=1.14$ for instance.
From this it is seen that the RY form does not give the proper
estimation of the degree of localization compared with the results
obtained in earlier works~\cite{Lowen,Das2}.

\begin{figure}
\epsfxsize=3.in
\centerline{\epsfbox{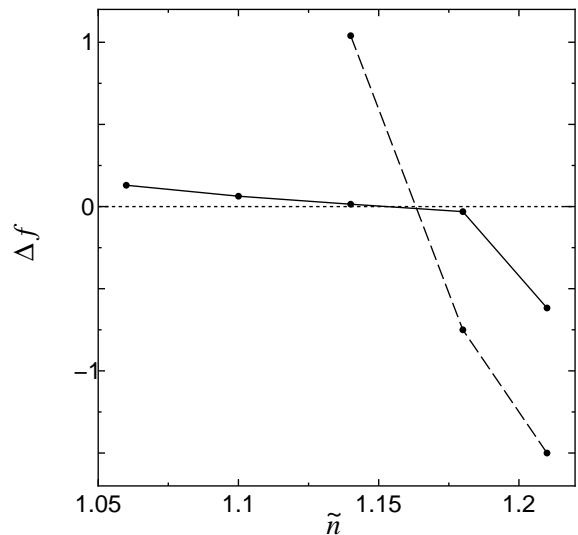}}
\caption{
Free energy differences $\Delta f$ of the weakly localize state (solid
line) and the highly localized state(dashed line)
as a function of density $\tilde n$.
}
\label{diff}
\end{figure}

In Fig.~\ref{diff}, we plotted the free energy differences $\Delta f$ of
the weakly and highly localized states as a function of the density
$\tilde n$.
From Fig.~\ref{diff}, we find that both the weakly and highly localized states
become more stable than
the uniform state at $\tilde n_{g,DFT} \simeq 1.15$, which is the
liquid-glass transition density from the energetics based on the DFT. 
Moreover, it is seen that the highly localized state is more
stable than the weakly localized state for higher densities.
In passing we note that our (first order) glass transition density 
$\tilde n_{g,DFT}= 1.15$ is rather close
to the one  $\tilde n_g=1.14$ found in Ref.~\cite{Singh}.

Finally, we compare our results from the energetics above with dynamical
informations supplied by our MD.
We find in Fig.~\ref{fs} that the intermediate scattering function
$F_s(q,t)$ begins to exhibit the two-step
relaxation at the density $\tilde n \simeq 1.06$, which 
corresponds precisely to the density where the 
free energy local minimum begins to appear in our DFT (see
Figs.~\ref{df}). 
Turning to the relaxation time $\tilde \tau$, we recall that the
density dependence of $\tilde \tau$ could be described by the power-law
$(\tilde n_{g,MD}-\tilde n)^{-\gamma}$ with $\tilde n_{g,MD} \simeq
1.15$.
This density happened to coincide with the density $n_{g,DFT}$ beyond
which the localized state is more stable than the uniform
liquid in the present DFT.
From these results, we consider that the DFT based energetics and 
dynamical behaviors related to slow dynamics are well correlated 
with each other.

In this paper, we reconsidered the DFT approach to the glass 
transition in the hard sphere system,
which was first undertaken by Singh {\it et al.}
We obtained hard sphere glasses by MD simulations without recourse to
the Bennett algorithm and the information on particle configurations
produced by the  MD simulations are used as 
input data when the free energy is calculated based on the DFT.
While only the uniform liquid state is stable at low density,
the free energy local minimum begins to appear at high
density $\tilde n \simeq 1.06$, where our MD shows that two-step
relaxation begins to appear.
This metastable glassy state becomes stable relative to uniform
liquid at $\tilde n_{g,MD}= 1.15$. 
Slow relaxation as represented by the $F_s(q,t)$ turned
out to be consistent with the energetics based on the DFT.

Before concluding this paper, we comment on recent developments in
studies of the DFT.
In recent years,
the so-called weighted density approximation (WDA)
for the free energy functional has been
developed~\cite{Tarazona,Ashcroft} and the modified version is also
introduced~\cite{Ashcroft2}.
Moreover, the fundamental measure theory (FMT) is
proposed~\cite{Rosenfeld} and gathering considerable interest.
It is well known that for highly localized states, such
methods are more accurate than the RY one.
This is because the former employs a non-perturbation
approximation whereas the latter relies a perturbation expansion
around the uniform state.
Several workers have already investigated the glass transition
by using the modified WDA method and found that the metastable
localized state is located at $\alpha \simeq 100$~\cite{Lowen,Das2}
in accordance also with our MD results .
Although the DFT based on the RY functional is still useful because of
its physical clarity, generality and simplicity,
we think that in view of the recent achievements
it is meaningful to employ the WDA or FMT method 
in order to obtain improved results.

Furthermore, we hope that the present DFT
approach will be applied to more complex systems.
As a model of a glass forming liquid, binary
Lennard-Jones~\cite{Kob} or soft-core system~\cite{Yamamoto} has
been investigated by large scale MD simulations.
Our approach can be readily applied to such system and would give new
insights into the glass transition from both thermodynamic and dynamical
viewpoints.


\end{multicols}

\begin{references}
\bibitem{units}
For all physical quantities, the following reduced units are used:
$\tilde r=r/\sigma$, $\tilde t=t\sqrt{{k_BT}/{m\sigma^2}}$, $\tilde
n=n\sigma^3 (n\equiv N/V)$, $\tilde q=q\sigma$, and $\tilde
\alpha=\alpha\sigma^2$.
\bibitem{numerical} It is remarked here that due to frequent 
collisions under compressed situation  a long time simulation at high
density is prohibitively difficult to perform.
\end{references}
\end{document}